\documentclass[prl,aps,superscriptaddress,showpacs]{revtex4}

\def \beq{\begin{equation}}         \def \eeq{\end{equation}}
\def \beqa{\begin{eqnarray}}        \def \eeqa{\end{eqnarray}}
\def \bea{\begin{array}}        \def \eea{\end{array}}

\usepackage{amsmath}
\usepackage{epsfig}
\usepackage[dvips]{color}

\begin{document}

%Title of paper
\title{Operator equality on entropy production in quantum Markovian master equations}
\author{Fei Liu}
\email[Email address:]{feiliu@buaa.edu.cn} \affiliation{School
of Physics and Nuclear Energy Engineering, Beihang University,
Beijing 100191, China}
\date{\today}

\begin{abstract}
{An operator equality on the entropy production for general
quantum Markovian master equations is derived without resorting
quantum stochastic trajectory and priori quantum definition of
entropy production. We find that, the equality can be still
interpreted as a consequence of time-reversal asymmetry of the
nonequilibrium processes of the systems. In contrast with the
classical case, however, the first order expansion of the
equality does not directly related to the mean entropy
production, which arises from noncommute property of operators
in quantum physics. }
\end{abstract}
 \maketitle

{\noindent\it Introduction} Irreversible processes can be seen
almost everywhere in nature. Imaging that a process of an
object falling into water starts from a static state and ends
with some position and velocity after a finite time interval.
However, if one wants to reverse the process by simply
reversing the object's velocity from the ending position, the
object never returns back its initial state after the same time
interval. This phenomenon obviously arises from energy
dissipation as heat due to friction between the object and its
reservoir. In modern thermodynamics, irreversible process is
always related to nonnegative entropy
production~\cite{Groot,Kondepudi}, or it is a manifestation of
the second law of thermodynamics. Although the law has been
rigidly established in macroscopic systems, in the past few
decades, interest in the entropy production or dissipated work
in small nonequilibrium systems has grown intensively due to
the finding of various fluctuation theorems or
relations~\cite{Bochkov77,Evans93,Gallavotti95,Kurchan98,Lebowitz99,Maes99,JarzynskiPRL97,
Crooks99,HatanoSasa01,Seifert05}. These remarkable relations
greatly deepen our understanding about the second law of
thermodynamics and nonequilibrium physics of small systems.
With the fluctuation relations clarified in the classical
systems, recently, we may see an trend of extending the
relations into quantum systems~\cite{Kurchan00,Yukawa00,
Tasaki00,DeRoeck04,Talkner07,Andrieux08,Crooks08,Esposito09,Campisi11,
Deffer11, Chetrite12,LiuF12}.

In this work, we present an {\it operator version} for the
entropy production equality in nonequilibrium systems that can
be described by quantum Markovian master equations. Because
Markovian description implies that the time change of external
sources does not affect reservoirs and is very slowly in
comparison with reservoir's relaxation time~\cite{Alicki79}, we
are not intended to state that the equality obtained here holds
even if the system is driven very far from equilibrium, e.g.,
like the entropy production equality derived by Deffer and
Lutz~\cite{Deffer11} using the sophisticated two-point energy
measurement statistics~\cite{Talkner07,Campisi11}. However, we
think that the price is worthwhile paying, since we can derive
an exact operator equality on the entropy production. Although
there have existed various quantum fluctuation relations in the
literature, to our knowledge, very few of them are in operator form~\cite{Chetrite12,LiuF12}.\\

{\noindent\it Time-reversal of system} We concern about the
irreversible process of an open quantum Markovian system $L_t$
during a time interval $(0,T)$. The equation of motion for the
system's density operator $\rho(t)$ is then
\begin{eqnarray}
\label{orgsystem}
\partial_{t}\rho(t)&=&L_{t}\rho(t)=L_{t}^{\rm irr}\rho(t)+L_{t}^{\rm irr}\rho(t),
\end{eqnarray}
where
\begin{eqnarray}
L_{t}^{\rm rev}\rho(t)=-i[H_t ,\rho(t)],
\end{eqnarray}
$H_t$ is free Hamiltonian of the system, and $L_{t}^{\rm irr}$
represents a dissipative term due to the interaction of the
system with a heat reservoir and it has a general
form~\cite{Davies74,Gorini76,Lindblad76,Breuer}
\begin{eqnarray} L^{\rm
irr}_t\rho(t)=\frac{1}{2}\sum_j{[ V_j,\rho(t) V^{\dag}_j]
+[V_j\rho(t), V^{\dag}_j]}.
\end{eqnarray}
Here we use the subscripts $t$ to indicate their possible
time-dependence except $V_j$ and $V^\dag_j$ for the simplicity
in notation. We define an alternative quantum Markovian system
$\tilde L_s$ as a time-reversal of the
system~(\ref{orgsystem}), if its density operator
$\tilde\rho(s)$ satisfies a master equation of
\begin{eqnarray}
\label{reversedsystem}
\partial_s \tilde\rho(s)=\tilde{L}_s\tilde\rho(s)=\tilde L^{\rm rev}_s\tilde\rho(s)+\tilde L^{\rm irr}_s\tilde\rho(s),
\end{eqnarray}
and
\begin{eqnarray}
\tilde{L}_s^{\rm rev}A&=&-\Theta L_{t'}^{\rm rev}[\Theta A\Theta^{-1}]\Theta^{-1}, \\
\tilde{L}_s^{\rm irr}A&=&\Theta L_{t'}^{\rm irr}[\Theta A\Theta^{-1}]\Theta^{-1},
\end{eqnarray}
where $A$ denotes an arbitrary operator, $t'$=$T$$-$$s$ is the
backward time~\cite{Kolmogorov31}, $\Theta$ is time-reversal
operator and we use a new time parameter $s$ (0$\le$s$\le$T)
for the time-reversal system. Specifically, if the system is
consistent with its time-reversal, we call such a kind of
system to be symmetric or invariable under time-reversal. These
definitions are in fact a simple quantum extension of those in
classical Markovian
process~\cite{Graham71,Risken72}.\\

\begin{figure}
\begin{center}
\includegraphics[width=0.8\columnwidth]{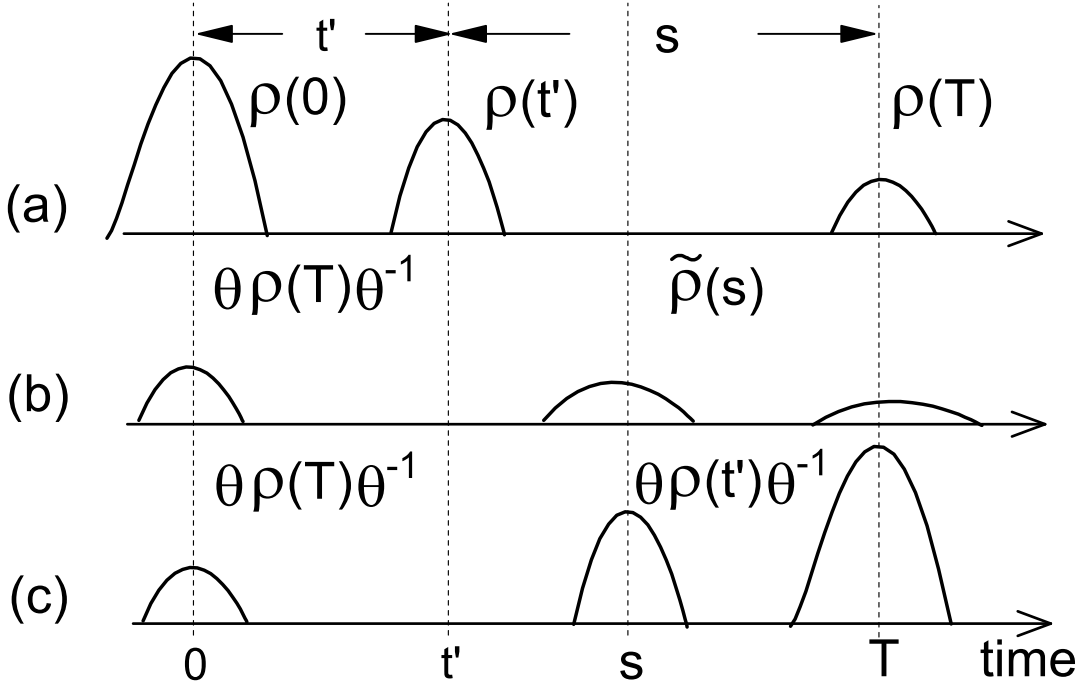}
\caption{(a) An irreversible time process of the system $L_t$.
(b)The real time process of the time-reversal system $\tilde{L}_s$
with a specified initial condition $\Theta\rho(T)\Theta^{-1}$.
(c) The imaginary process constructed by the time reversed operator $\Theta \rho(t')\Theta^{-1}$,
which is usually inconsistent with $\tilde\rho(s)$ due to dissipation.} \label{figure1}
\end{center}
\end{figure}

{\noindent\it Operator $R(t',T)$} If the dissipation term
$L^{\rm irr}_t$ vanishes, and the Hamiltonian $H_t$ is even
under the time reversal, namely, $H_s$=$\Theta
H_{t'}\Theta^{-1}$, one may easily prove that a time-reversed
operator, $\Theta \rho(t')\Theta^{-1}$, is the solution of the
time-reversal system~(\ref{reversedsystem}) with a specified
initial condition $\Theta \rho(T)\Theta^{-1}$. We call such
kind of solution to be time-reversible. In addition, if the
system is symmetric under time-reversal and is at thermal
equilibrium state $\rho_0$, the state is also time-reversible
and invariable specifically, i.e.,
$\Theta\rho_0\Theta^{-1}$=$\rho_0$. Generally speaking, as open
quantum system has a dissipative term, if it is perturbed by
time-dependent sources, and/or is relaxing to its equilibrium
state, the solution $\rho(t)$ is no longer reversible. This
observation could be simply quantified if we introduce an
operator $R(t',T)$ satisfying a relationship of
\begin{eqnarray}
\label{Roperator}
\tilde{\rho}(s)=\Theta R(t',T)\rho(t')\Theta^{-1},
\end{eqnarray}
and $R(T,T)$$=$$1$. Obviously, if the solution $\rho(t)$ was
reversible, $R(t',T)$ would equal the identity operator during
the whole time interval; otherwise it would not.
Fig.~\ref{figure1} schematically explains the reason which we
define the operator $R(t',T)$.

Substituting (\ref{Roperator}) into Eq.~(\ref{orgsystem}), we
can obtain an equation of motion for $R(T,t')$ with respect to
the backward time $t'$ given by
\begin{eqnarray}
\label{EOMR}
\partial_{t'} R(t',T)&=&-L^{*}_{t'} R(t',T) - R(t',T)[\partial_{t'}{\rho(t')} - L_{t'}{\rho(t')}]{\rho(t')}^{-1}\\ \nonumber
&&-\{\sum_j [R(t',T)V_j \rho(t'),V_j^\dag]-[R(t,T)\rho(t')V_j^\dag,V_j]\}\rho(t')^{-1}\\
&=&-L^{*}_{t'} R(t',T)-{\cal O}_{t'} R(t',T),\nonumber
\end{eqnarray}
where the adjoint generator is $L^*_{t'}A$$=$$i[H_{t'},A]$$+$$
(1/2)\sum_j [V_j^{\dag},A]V_j$$ +$$ V^\dag_j[A,V_j]$ and
$L^{*}_{t'}$$1$=0 particularly. In the remaining part, we use
the superscript $(\cdots)^{\star}$ to denote an adjoint
super-operator with respect to the trace unless otherwise
stated. Notice that this is a terminal condition problem
instead of conventional initial condition problem. If we regard
the term having the super-operator ${\cal O}_t'$ in
Eq.~(\ref{EOMR}) as a perturbation, which is rational because
we concern about the deviation of $R(t',T)$ with respect to
$1$, we can obtain a formal solution for $R(t',R)$ using the
Dyson series~\cite{Chetrite12,LiuF12}:
\begin{eqnarray}
\label{DysonexpansionR}
R(t',T)=[G^\star(t',T)+\sum_{n=1}^{\infty}\int_{t'}^Tdt_1\cdots\int_{t_{n-1}}^T dt_n\prod_{i=1}^n
G^\star(t_{i-1},t_i){\cal O}_{t_i} G^\star(t_n,T)] R(T,T)
\end{eqnarray}
where $G^\star(t',T)$$=$${\cal T}_{+}\exp [\int_{t'}^T d\tau
L^*_{\tau}$] is the adjoint propagator of the system, and
${\cal T}_{+}$ denotes the antichronological
time-ordering operator~\cite{Breuer}.\\

\noindent{\it operator equality on the entropy production.}
Although we have the formal expression for $R(t',T)$, it is not
satisfied because the perturbation in
Eq.~(\ref{DysonexpansionR}) involves $\rho(t')$, which makes
its physical explanation ambiguous. In order to proceed
further, in this work we restrict the system to those
satisfying instant detailed balance condition with respect
their instant thermal equilibrium state
$\rho_0(t)$~\cite{Alicki76,Kossakowski77}:
\begin{eqnarray}
&&L^{\rm rev}_t\rho_0(t)=0,\\
&&L^{\rm irr}_t A\rho_0(t)={L^{\rm irr\star}_t}[A]\rho_0(t).
\end{eqnarray}
These conditions mean that the system always relaxes to its
thermal equilibrium state $\rho_0(t)$, if the external source
is fixed at the value with time point $t$~\cite{Spohn78EP}.
This limitation seems not very strict from physical point of
view. Under this circumstance, we may define an auxiliary
operator $R_0(t',T)$ as follows:
\begin{eqnarray}
\label{Eqsplitting}
R(t',T)\rho(t')=R_0(t',T)\rho_0(t').
\end{eqnarray}
One can prove that the equation of motion for $R_0(t',T)$ is
analogous to Eq.~(\ref{EOMR}) except that the previous
perturbation term is now replaced with
\begin{eqnarray}
\label{worksuperoperator}
-{\cal W}_{t'} R_0(t',T)%&=&{\cal W}_{t'} R_0(t',T)
=-R_0(t',T)\partial_{t'} \rho_0(t')\rho_0(t')^{-1},
%{\cal Q}_{t'} R_c(t',T)%&=&{\cal Q}_{t'} R_c(t',T)
%&=&\sum_j V_j R_c(t',T) V^\dag_j-V^\dag_j R_c(t',T) V_j,
\end{eqnarray}
and the terminal condition of $R_0(t',T)$ is
$\rho(T)\rho_0^{-1}(T)$. The reader is reminded that the new
defined ${\cal W}_t$ is  a super-operator, thought its action
on an operator is a simple multiplication from the operator's
right-hand side.

As we mentioned previously, a deviation of $R(t',T)$ with
respect to 1 indicates the time-irreversibility of the solution
$\rho(t)$. Hence, selecting $t'$=0, regarding the logarithms of
the all density operators as ``small" operator and using the
relation~(\ref{worksuperoperator}) and the Dyson series for
$R_0(0,T)$, we can expand $R(0,T)$ around 1 until the first
order:
\begin{eqnarray}
\label{1storddef}
R(0,T)&=& 1 +  \Theta^{-1}\ln \tilde{\rho}(T)\Theta -\ln\rho(0) +\cdots\\
\label{1stordwork}
&=& 1 + G^\star(0,T)\ln\rho(T) -G^\star(0,T)\ln\rho_0(T)+
\int_0^T d\tau G^\star(0,\tau)\partial_\tau\ln\rho_0(\tau)+\ln\rho_0(0)-\ln\rho(0)+\cdots%\\
\label{1stordheat}
\end{eqnarray}
The first equation arises from the
definition~(\ref{Roperator}), and in the latter equation we
have used a simple property of $G^\star(0,T)1=1$. We must point
out explicitly that the first order expansion has also been
applied to the super-operator ${\cal W}_\tau$. Notice that the
above equation also holds if the initial time 0 is replaced
with arbitrary time point $t'$ ($\le T$).

We find that an equal of Eqs.~(\ref{1storddef}) and
~(\ref{1stordwork}) provides us an microscopic expression for
the second law of thermodynamics. To see it, Multiplying them
with $\rho(0)$ and taking the trace, we have
\begin{eqnarray}
\label{meanEP}
&&\langle \ln \rho(0)\rangle_0-\langle
\Theta^{-1}\ln\tilde\rho(T)\Theta\rangle_0\nonumber\\
=&&[-\langle
\ln\rho(T)\rangle_T+\langle\ln\rho(0)\rangle_0]+[\langle \ln
\rho_0(T)\rangle_T-\langle \ln \rho_0(0)\rangle_0 - \int_0^T d\tau\langle \partial_\tau\ln\rho_0(\tau)\rangle_\tau]
\end{eqnarray}
where $\langle A\rangle_\tau$$=$${\rm Tr}[A\rho(\tau)]$, we
have used properties of ${\rm Tr}[G^\star(t',T)(A)B]={\rm Tr}[A
G(T,t')(B)]$ and $G(\tau,0)\rho(0)$=$\rho(\tau)$. Here
$G(T,t')$ is the system's propagator and equals ${\cal
T}_{-}\exp[ \int_{t'}^T d\tau L_{\tau}]$, and ${\cal T}_{-}$ is
the chronological time-ordering operator~\cite{Breuer}. We see
that, on the right-hand side of Eq.~(\ref{meanEP}), the terms
in the first square bracket is the change of von Neumannn
entropy of the system,
$S(\rho(\tau))$=$-\langle\ln\rho(\tau)\rangle_\tau$, while the
terms in second square bracket is the mean heat transfer from
the system to the heat reservoir. The latter is consequence of
the first law of thermodynamics~\cite{Alicki79}, which becomes
obvious if the quantum master equation is obtained under the
weak coupling limits and the equilibrium state $\rho_0(t)$ has
a canonical ensemble~\cite{Spohn78,Gorini78,Davies74}.
According to the principles of phenomenological
thermodynamics~\cite{Groot}, the whole expression on the
right-hand side is no other than the mean entropy production of
the irreversible quantum process and it is always assumed to be
nonnegative. Deffer and Lutz obtained the same expression by
the principle directly~\cite{Deffer11}. Compared with their
argument, here two interesting features are revealed by the
terms on the left-hand side of Eq.~(\ref{meanEP}). First,
because of the equal between two sides, we have a new form for
the mean entropy production, which is based on the initial and
terminal density operators of the original and time-reversal
systems, respectively. In particular, if the initial state is
time-reversal invariable, e.g., the system initially being a
thermal equilibrium, the expression of left-hand side is just a
quantum relative entropy,
$S(\rho(0)||\tilde\rho(T))$~\cite{Cover}. Under this
circumstance, the nonnegative property of the mean entropy
production has a rigors mathematical foundation rather than
phenomenological reason. Second, if we concerned about a
relaxation process of a time-reversal symmetric system from
nonequilibrium initial state $\rho(0)$ to the thermal
equilibrium $\rho_0$ after time $T$ without any external
perturbation~\cite{Spohn78EP}, the expression on the left-hand
side would become $S(\rho(0)||\rho_0)$. The reason is that the
time-reversal system starts from the equilibrium initial state
$\rho_0$ that is also the terminal state of the relaxation
process. Therefor, $\tilde\rho(T)$ would equal $\rho_0$. In
this case, the equal between two sides of Eq.~(\ref{meanEP})
becomes trivial.

Now we present an operator equality on the entropy production:
\begin{eqnarray}
\label{EPequality}
1&=&\langle \rho(T)\rho_0(T)^{-1}{\cal T}_{-}\exp[\int_0^T d\tau \partial_{\tau}\rho_0(\tau)\rho_0^{-1}(\tau)]
\rho_0(0)\rho^{-1}(0)\rangle
\end{eqnarray}
Proof:
\begin{eqnarray}
1&=&{\rm Tr}[ \Theta^{-1}\tilde\rho(T)\Theta]={\rm Tr}[R_0(0,T)\rho_0(0)]\nonumber\\
&=&{\rm Tr}\{[G^\star(0,T)+\sum_{n=1}^{\infty}\int_{t'}^Tdt_1\cdots\int_{t_{n-1}}^T dt_n\prod_{i=1}^n
G^\star(t_{i-1},t_i){\cal W}_{t_i} G^\star(t_n,T)] [\rho(T)\rho_0^{-1}(T)]\rho_0(0) \}\nonumber\\
&=&{\rm Tr}[\rho(T)\rho^{-1}_0(T)G(T,0)\rho_0(0)]+\int_0^Tdt_1{\rm Tr}[\rho(T)\rho^{-1}_0(T)
G(T,t_1) \partial_{t_1}\rho_0(t_1)\rho^{-1}_0(t_1)G(t_1,0)\rho_0(0)]+\cdots\nonumber\\
&=&\langle\rho(T)\rho_0(T)^{-1}\rho_0(0)\rho^{-1}(0)\rangle + \int_0^Tdt_1\langle \rho(T)\rho_0(T)^{-1}
\partial_{t_1}\rho_0(t_1)\rho_0^{-1}(t_1) \rho_0(0)\rho^{-1}(0)\rangle+\cdots\nonumber
%&=&\langle \rho(T)\rho_0(T)^{-1}{\cal T}_{-}\exp[\int_0^T d\tau \partial_{\tau}\rho_0(\tau)\rho_0^{-1}(\tau)]
%\rho_0(0)\rho^{-1}(0)\rangle,
\end{eqnarray}
The transformation from the third line to the forth line is
based on the definition of multi-time correlation for operators
in quantum master equations~\cite{Chetrite12,GardinerQM}. We
must emphasize that, expanding the time-ordered exponential
term in the operator equality to the first order does not
simply lead into the mean entropy production
equation~(\ref{meanEP}), since
$\partial_\tau\rho_0(\tau)\rho_0^{-1}(\tau)$ usually does not
equal $\partial_\tau \ln\rho_0(\tau)$, which is unique only in
quantum physics.\\

{\noindent\it Quantum Jarzynski equality} Chetrite and Mallick
have derived an operator Jarzynski equality using a modified
dynamics for the accompanying density matrix~\cite{Chetrite12}.
Here we can give an alternative derivation using the same
sprint deriving Eqs.~(\ref{EPequality}). Following the
conventions of proving Jarzynski
equality~\cite{JarzynskiPRL97,Crooks99,Chetrite12}, we assume
that the system has instant equilibrium solutions $\rho_0(t)$
satisfying the detailed balance condition, and the system is
initially in equilibrium of $\rho_0(0)$. Analogous to previous
case, we are still interested in comparing two processes from
the original system and the time-reversal system
using~(\ref{Roperator}), but here the initial density operator
of the latter process is replaced with
$\tilde{\rho}(0)$$=$$\rho_0(T)$. Therefor, the terminal
condition $R(T,T)$ becomes $\rho_0(T)\rho^{-1}(T)$ instead of
previous 1. Because of the instant detailed balance condition,
we may introduce an auxiliary $R_0(t',T)$
as~(\ref{Eqsplitting}) again and doing the same calculation to
obtain the operator Jarzynski equality
\begin{eqnarray}
\label{JEoperator}
1=\langle {\cal T}_{-}\exp[\int_0^T d\tau \partial_{\tau} \rho_0(\tau)\rho_0^{-1}(\tau)]\hspace{0.1cm}\rangle_0
\end{eqnarray}
Here the subscript $0$ is to indicate that the equality holds
only for equilibrium initial condition. Obviously£¬ the
operator equalities~(\ref{EPequality}) and~(\ref{JEoperator})
are not the same unless both the initial and terminal states of the nonequilibrium process
are thermal equilibriums. \\

{\noindent\it Discussion and conclusion} By investigating the
difference between the density operators of quantum Markovian
master equation and its time-reversal, in this work we present
an operator equality on the entropy production. Our discussion
is based on there key assumptions. The first two assumptions
are a description of perturbed quantum system using the
Markovian master equation and the system satisfying instant
detailed balance condition, respectively. Although the two
assumptions seem to limit the validity of the operator equality
in very far from equilibrium regime, we should emphasize that
analogous assumptions in fact have been implied in derivation
of the various fluctuation relations in classical Markovian
systems. The last assumption is the existence of $\ln\rho(t)$,
or equivalently the system's density operator to be invertible.
So far, we do not find a satisfying mathematical or physical
approach to justify it. Hence, we have to leave it for future
study. Finally, we may point out that an extension of present
theory to the classical Markovian processes is very straightforward.\\

{\noindent We appreciate Prof. Chetrite for sending their
inspiring work~\cite{Chetrite12} to us. This work was supported
by the National Science Foundation of China under Grant No.
11174025.}

%&\approx& 1+ G^\star(0,T)\ln\rho(T)+\int d\tau
%G^\star(0,\tau){\cal Q}_\tau(1)-\ln\rho(0)

%That is, the effect of $\rho(t')$ has
%been put into terminal conditions instead of the whole
%equations of motion.

%\begin{eqnarray}
%\label{firstlaw} \langle \ln \rho_0(0)\rangle_0 -\langle \ln
%\rho_0(T)\rangle_T=-\int_0^T d\tau\langle {\cal
%W}_\tau(1)\rangle_\tau + \int_0^Td\tau \langle {\cal
%Q}_\tau(1)\rangle_\tau
%\end{eqnarray}

\end{document}